\documentclass[]{aa}
\usepackage{natbib} 
\usepackage{graphicx}
\usepackage{txfonts}

\newcommand{\Teff}  {\mbox{T$_\mathrm{eff}$}\,}
\newcommand{\FeH}   {\mbox{[Fe/H]\,}}
\newcommand{\logg}  {\mbox{$\log$ g}\,}
\newcommand{\tgm}   {\mbox{(\Teff, \logg, [Fe/H])} \,}

\begin{document}

\title{The PASTEL catalogue of stellar parameters\thanks{The catalogue can be queried through a dedicated web interface at http://pastel.obs.u-bordeaux1.fr/. It is also available in electronic form at the Centre de Donn\'ees Stellaires in
Strasbourg  (http://vizier.u-strasbg.fr/viz-bin/VizieR?-source=B/pastel)}}
\titlerunning{PASTEL catalogue}

\author{Soubiran C.\inst{1}, Le Campion J.-F.\inst{1}, Cayrel de Strobel G.\inst{2}\and  Caillo A.\inst{3}}

\offprints{C. Soubiran, \email{soubiran@obs.u-bordeaux1.fr}}

\institute{Laboratoire d'Astrophysique de Bordeaux (LAB-UMR 5804), CNRS, Universit\'e Bordeaux 1,
2 rue de l'Observatoire, BP 89, F-33271 Floirac cedex
\and
GEPI, Observatoire de Paris, CNRS, Universit\'e Paris Diderot, Place Jules Janssen, 92190 Meudon, France 
 \and
Observatoire Aquitain des Sciences de l'Univers (OASU-UMS 2567), CNRS, Universit\'e Bordeaux 1, 
2 rue de l'Observatoire, BP 89, F-33271 Floirac cedex
 }
\date{Received  / Accepted}

\abstract 
{} 
{The PASTEL catalogue is an update of the \FeH catalogue, published in 1997 and 2001. It is a bibliographical compilation of stellar atmospheric parameters providing \tgm determinations 
obtained from the analysis of high resolution, high signal-to-noise 
spectra, carried out with model atmospheres. PASTEL also provides determinations of the one parameter \Teff based on various methods. It is aimed in the future to provide also homogenized atmospheric parameters and  elemental abundances, radial and rotational velocities. A web interface has been created to query the catalogue on elaborated criteria. PASTEL is also distributed through the CDS database and VizieR.}
{To make it as complete as possible, the main journals have been surveyed, as well as the CDS database, to find relevant publications. The catalogue is regularly updated with new determinations found in the literature.}
{As of Febuary 2010, PASTEL includes 30151 determinations of either  \Teff or \tgm for 16649 different stars corresponding to 865 bibliographical references. Nearly 6000 stars have a determination of the three parameters \tgm with a high quality spectroscopic metallicity.
}
{}

		
\maketitle
		

\section{Introduction}

The knowledge of stellar atmospheric parameters \tgm is mandatory to estimate the abundance of any chemical element in a stellar atmosphere, which then traces either the composition of the 
interstellar medium from which the star formed or the various physical processes which may alter the initial composition.  Atmospheric parameters are thus essential in many reasearch areas related to the physics of stars and galaxies. 

The detailed analysis of high resolution, high signal-to-noise spectra is the only primary method to estimate the chemical composition of stellar photospheres, from which other indirect methods of metallicity determinations can be calibrated. The general approach of a spectroscopic detailed analysis is to derive
the iron abundance by
matching equivalent widths of weak lines or spectral intervals of an observed spectrum to those computed 
from a grid of model
atmospheres of various effective temperatures, gravities and metallicities. The higher the resolution and signal-to-noise ratio, the better are supposedly the results. However, each individual study has its own observational characteristics, preferred stellar atmosphere models, atomic line data and analysis methods. The consequence is a lack of homogeneity in the results from one study to the next which makes it very cumbersome to combine the determinations. Defining a common scale for atmospheric parameters is a mandatory but difficult task, which starts with the compilation of the best studies available in the literature. This is the aim of the PASTEL catalogue, which follows the previous \FeH catalogue.

The publication of the \FeH catalogue started with a first version in 1980 \citep{cay80}, continued with three other versions \citep{cay81,cay85, cay92} superseded by the 1996 edition \citep{cay97}. Then the content of the 2001 edition \citep{cay01} was modified by limiting the temperature range to stars cooler than 7000K and by removing references older than 1980. It was time to update the \FeH catalogue. Since 2001, spectroscopic observations of individual stars at high resolution and high signal-to-noise (S/N) have been intensive. Several research topics are particularly productive regarding the number of analysed stars. Studies on the chemical composition of stars with planets, on the metallicity distribution and gradients in the Galaxy and on the chemical composition of metal-poor stars have produced extensive lists of stars, based on high-quality data, with a good fraction of stars which have been observed at high spectral resolution for the first time. 

The methods have evolved. Due to the growing number of spectra to be processed, it becomes more frequent that spectroscopic analyses are automatised. It can be either the measurement of equivalent widths, or the fitting of individual lines or spectral regions to computed ones. As a consequence, recent papers tend to present larger number of stars than before. A typical example of such an extensive paper is the one by \citet{Val05} presenting a uniform catalogue of stellar properties for 1040 nearby F, G, and K stars.

Effective temperature is a critical parameter in spectroscopic analyses, because errors in \Teff lead to significant errors in measured abundances. Several recent studies have contributed to a significant increase of the number of stars with precise determinations of \Teff. We have included these studies in PASTEL, even if they are not based on high-resolution spectra. 

In this new catalogue we come back to the content of the 1996 edition of the \FeH catalogue including hot stars, and references older than 1980. We complete it with the 2001 edition and with new references gathered from the literature. Thus, PASTEL supersedes the two previous versions of the \FeH catalogue. In the two latest \FeH catalogues, field and cluster stars were presented in separate tables. Here we have them in a single catalogue.

A webserver has been created for a convenient query of the catalogue. It can be queried by a list of identifiers resolved by Simbad, or by contraints on the atmospheric parameters, or on equatorial coordinates, B and V or 2MASS J, H, K magnitudes, as well as by name of authors, year of publication or bibcode. 

The numbers given in this paper are those corresponding to the version of PASTEL in Febuary 2010. Because the catalogue is updated regularly, these numbers are changing accordingly.

\section{Description of the catalogue}

\subsection{Identification and basic data}
Only identifiers of stars resolved by Simbad are considered in PASTEL. As a consequence, the peculiar case of spectroscopic binaries in which the components could be resolved at high resolution and studied separetely are not included in the catalogue. We will consider these cases in future versions. 

We have adopted a rule to identify a star with a single name in the catalogue. We choose in priority its HD number, if that is not available we take by order :  BD, CD, HIP, LHS, NLTT, LTT, CPD. If a star has none of these identifiers, we adopt the one given in the publication if resolved by Simbad. 

For each star entered in PASTEL, Simbad is queried to retreive its equatorial coordinates and B, V,  J, H, K magnitudes. The catalogue can thus be searched either by zone in the sky or by magnitude interval.

\subsection{Effective temperature}
The direct approach to compute the effective temperature of a star,  from its angular diameter and total flux at Earth, is only applicable for a limited number of nearby stars. There are other indirect or semi-empirical methods using continuum and spectral-line sensitivities to temperature. In practice, there are several ways to implement these methods, each implementation giving its own temperature scale. As a consequence, the comparison of \Teff from one author to another may show some systematic differences. The most significant publications which have provided extensive lists of \Teff have been included in PASTEL. They are presented in Table \ref{tab:teff}. 

\begin{table*}
\centering
\caption{Major publications, in terms of number of stars,  providing \Teff determinations.}
\label{tab:teff}
\begin{tabular}{@{}l | r | c | l @{}}
\hline 
Reference & \# stars & \Teff range & Method \cr
\hline 
\citet{Alo96, Alo99} &    753 & 3562 - 9232  & IRFM \cr
\citet{Bla98} &   420  & 4063 - 9720 & IRFM \cr
\citet{diB98} &   533  & 3846 - 10021 & surface brightness + V-K calibration\cr
\citet{Kov03, Kov04, Kov06} & 645 & 3700 - 6118 & Line Depth Ratios \cr
\citet{Mas06} & 10999  & 3748 - 7987  & SED fit + V-K calibration\cr
\citet{Ram05} &  754 & 3535 - 9233 & IRFM \cr
\citet{Gonz09} & 814 & 3745 - 7489 & IRFM  + colour calibrations\cr
\hline
\end{tabular}
\end{table*}

\begin{figure}[t]
\resizebox{\hsize}{!}{\includegraphics{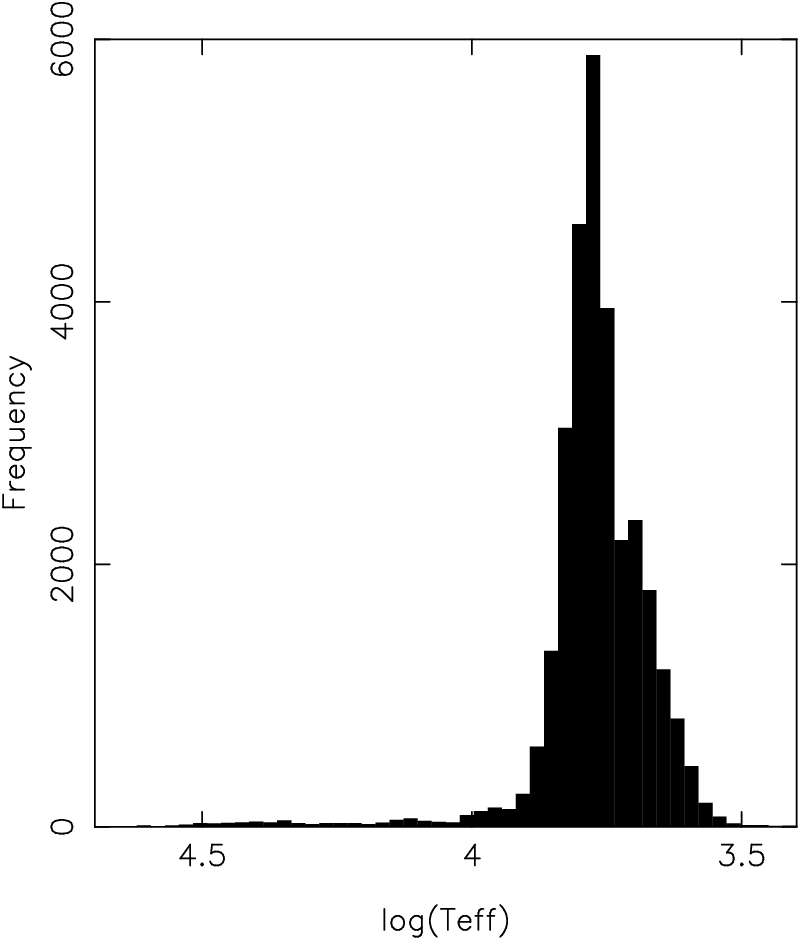}}
\caption{Histogram of all effective temperature determinations in PASTEL. }
\label{Teff_histo}
\end{figure}

The histogram of all \Teff in PASTEL is shown in Fig.~\ref{Teff_histo}. The vast majority of \Teff determinations are in the FGK regime. 

If quoted in the articles, individual errors on \Teff are given in PASTEL. On average these errors, which are internal, are at the level of  1.1\%. This can be compared to the dispersion around the mean of \Teff determinations for the fraction (24\%) of stars in PASTEL which have at least two available values. On average this dispersion is 1.3\% and reflects both internal and external errors, including differences in temperature scales. We have not attempted to make a more detailed analysis of the different temperature scales because the \Teff  determinations that we compare are not independant. Most studies use common reference data, even if the calibration methods are different. 

\subsection{Logarithm of surface gravity, \logg}
The surface gravity of a star is directly given by its mass and radius. It is a measure of the photospheric pressure of the stellar atmosphere. A direct measurement is possible from eclipsing spectroscopic binaries. 
Popular indirect methods use the ionization balance of iron in which \logg is tuned until the metallicity obtained from the FeI and FeII lines agrees, or wings of strong lines,  broadened by collisional damping. For nearby stars, the parallax is often used once the effective temperature is determined. When available, errors on \logg  are on average around 0.10 dex.

\subsection{Metallicity, \FeH} 
As usual \FeH is defined by \\

\FeH = $\log$(Fe/H)$_{\rm star} - \log$(Fe/H)$_{\rm Sun}$,\\

\noindent where Fe/H is the ratio of the number of iron atoms to the number of hydrogen atoms
in the atmosphere of either the star and the Sun. 

For this bibliographical compilation, we have only gathered \FeH determinations based on high resolution, high signal-to-noise spectra. We have considered that the high spectral resolution is above R=30\,000. In general high signal-to-noise implies a ratio higher than 100, although it is possible that some results included in PASTEL have been obtained on slightly lower S/N. 

\begin{figure}[t]
\resizebox{\hsize}{!}{\includegraphics{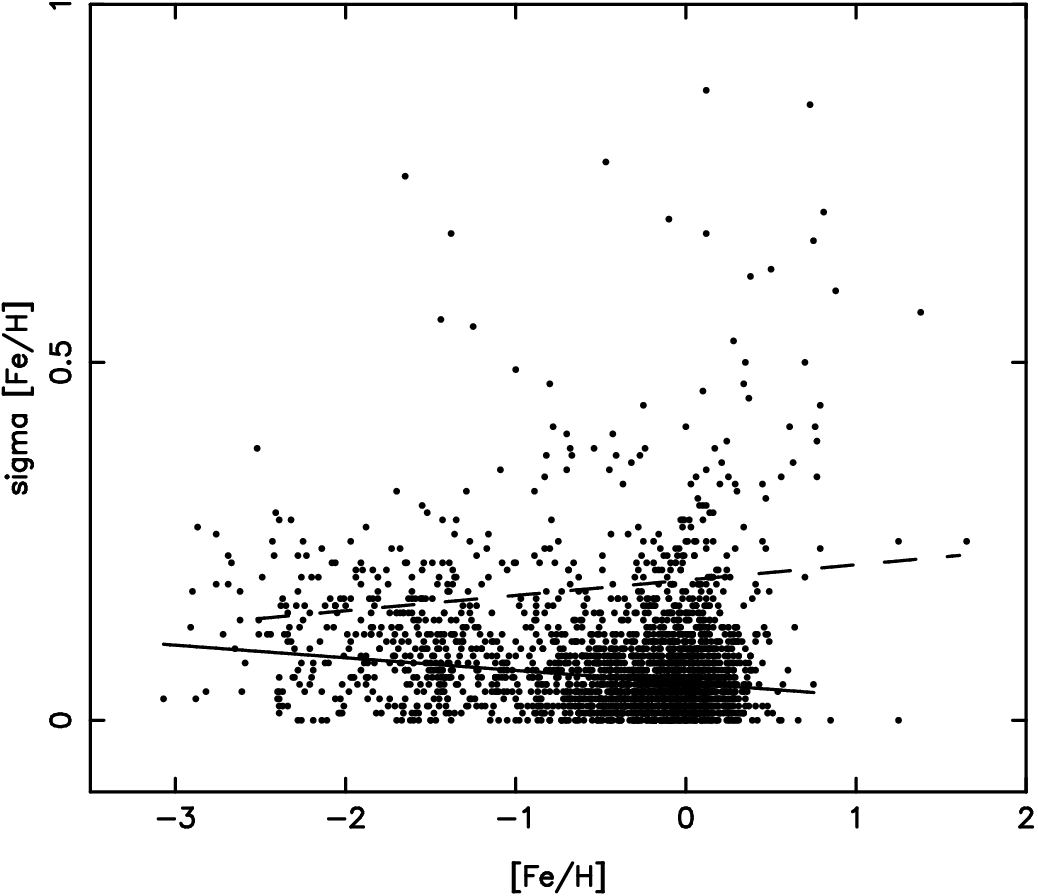}}
\caption{Dispersion of \FeH estimations as a function of  mean \FeH for 2731 stars with at least two available determinations. The dashed line corresponds to the linear fit for stars hotter than 7000K, while the continuous line corresponds to FGK stars and references posterior to 1990. }
\label{FsF}
\end{figure}

Here again it is worthwhile comparing the typical internal errors quoted in the publications, 0.06 dex on average, to the real dispersion of \FeH determinations when several of them are available for a given star. Among the stars which have \FeH determined, 2731 (44\%) have at least two determinations available with a typical dispersion of 0.08 dex. The dispersion is however related to the metallicity regime and the temperature as shown in Fig.~\ref{FsF}. For stars more metal-rich than \FeH=-1.0, the dispersion is essentially below 0.1 dex, except for stars hotter than 7000K, which sometimes exhibit very discrepant metallicity determinations. For metal-poor stars the dispersion is essentially below 0.2 dex. The situation is however slightly improved when the oldest determinations are not considered. 

We have investigated the differences of metallicity scales in more detail by comparing studies with at least 100 stars in common. Table \ref{t:comp_feh} gives the simple statistics, with a $3\sigma$ rejection, of two by two comparisons involving eight extensive and homogeneous datasets. 
Each homogeneous set is designated by the name of the first author and may include several papers based on the same model atmospheres, line data and analysis methods :
\begin{itemize}
 \item Fuhrmann = \citet{Fuh98c, Fuh98a, Fuh98b, Fuh04, Fuh08}, 300 stars
 \item Gratton = \citet{Gra96,Gra03}, 398 stars
 \item Hekker = \citet{Hek07}, 366 stars
 \item Luck = \citet{Luc06,Luc07}, 222 stars
 \item McWilliam = \citet{McW90}, 671 stars 
 \item Mishenina = \citet{Mish01,Mish04,Mish06,Mish08}, 557 stars
 \item Ram{\'{\i}}rez = \citet{Ram07}, 523 stars
 \item Valenti = \citet{Val05} , 1040 stars 
\end{itemize}

It is clear from these results that even though each of these studies can be considered of very high quality, none of them perfectly agrees with the others. Either there is an offset between the zero point of the metallicities, or the dispersion is significant. The best agreement, i.e.  smallest offset and dispersion, is achieved between Fuhrmann's datasets and the large sample of \cite{Val05}. Stars in common are however mainly G dwarfs of solar metallicity. The largest offset and dispersion are obtained when comparing the metallicities of giants studied by \citet{Hek07} and \citet{McW90}. A large dispersion is also seen in the comparison of Gratton's and Mishenina's datasets, which have a significant fraction of metal-poor stars in common. It is also worthwhile noting that the metallicity scales of  \citet{Ram07} and \citet{Val05}, the two largest samples of homegeneous stellar parameters for nearby disk stars, are shifted by 0.07 dex. This demonstrates the non-homogeneity of spectroscopic metallicities, supposed to be of the best quality.

\begin{table}
\centering
\caption{Two by two comparison of homogeneous sets of spectroscopic metallicities (see text). A simple statistic with  $3\sigma$ rejection has been applied. $N_c$ is the number of common stars, $N_o$ is the number of $3\sigma$ outliers.}
\label{t:comp_feh}
\begin{tabular}{l | l | r | r | r | r }
\hline 
Reference 1 & Reference 2 & $\Delta_{\FeH}$ & $\sigma_{\FeH}$ & $N_c$  &   $N_o$ \cr
\hline 
Fuhrmann        & Mishenina       &  0.01 &  0.07 & 137 &   3 \cr 
Fuhrmann        & Valenti         & -0.03 &  0.05 & 179 &   3 \cr 
Gratton         & Mishenina       & -0.02 &  0.10 & 103 &   0 \cr 
Gratton         & Ram{\'{\i}}rez   &  0.02 &  0.08 & 110 &   2 \cr 
Hekker          & McWilliam       &  0.07 &  0.10 & 211 &   5 \cr 
Luck            & Valenti         & -0.01 &  0.06 & 136 &   4 \cr 
Mishenina       & Valenti         & -0.04 &  0.07 & 162 &   4 \cr 
Ram{\'{\i}}rez   & Valenti         & -0.07 &  0.06 & 137 &   3 \cr 
\hline
\end{tabular}
\end{table}

\subsection{Bibliographical references}
The tables of content of the main astronomical journals are regularly surveyed to search for relevant publications, with data available in numerical form. VizieR tables at the CDS are also checked for new entries, through a query by Unified Content Descriptor equal to PHYS\_ABUND\_FE/H and PHYS\_TEMP\_EFFEC. For each publication introduced in PASTEL we give the name of the first author, the year of publication and the bibcode for an easy retrieval and citation of the corresponding article.

The most substantial contribution to the catalogue comes from the A\&A journal with nearly 21000 entries in PASTEL. Then ApJ and ApJS are quoted $\sim$5000 times, followed by MNRAS with 1383 occurences.

Although PASTEL is intended to be exhaustive, the lack of manpower does not allow us to be complete. We recommend users of the catalogue to notify us of missing references which should be included in the catalogue.

\section{Stellar content of the catalogue}

The sample of stars in PASTEL
cannot be considered as
representative of the stellar content of the solar neighbourhood. 
 Obviously, the various observing programs, dealing with very different astrophysical problems, from 
which the catalogue 
was built, introduce some biases in the distributions of \tgm.

\begin{figure}[t]
\resizebox{\hsize}{!}{\includegraphics{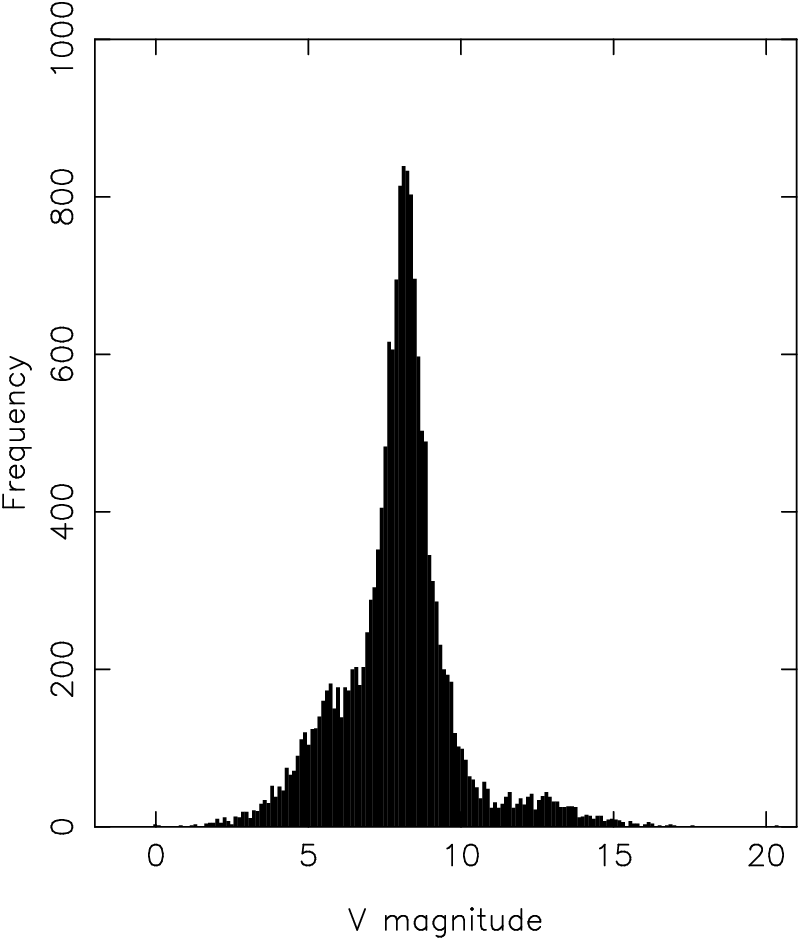}}
\caption{Histogram of the V magnitude in PASTEL, available for 16594 stars. }
\label{V_histo}
\end{figure}

In Febuary 2010 PASTEL included 16649 different stars. Their histogram in V magnitude is presented in Fig.~\ref{V_histo}. Although the situation of faint stars is improving, 90\% of the stars in PASTEL are still brighter than V=9.75. There are less than 50 stars fainter than V=15.5. 

\begin{figure}[t]
\resizebox{\hsize}{!}{\includegraphics{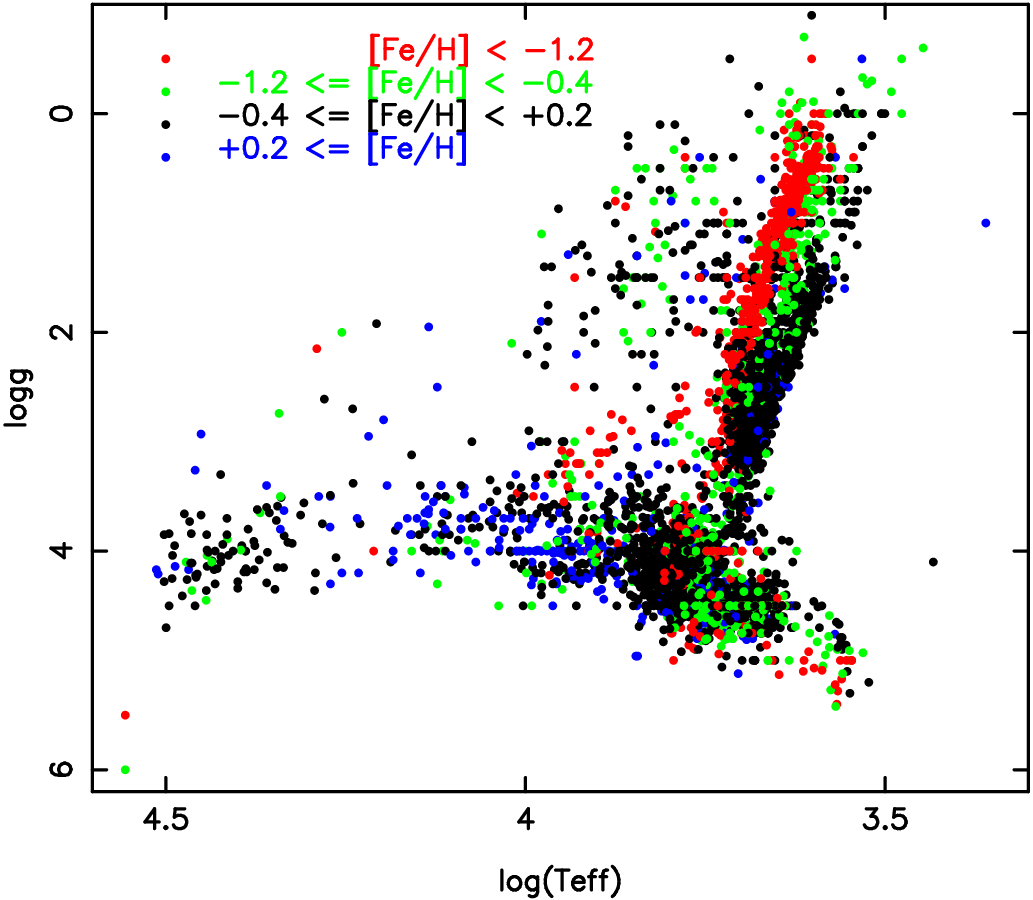}}
\caption{log(\Teff) vs. \logg in several regimes of metallicity for nearly 6000 different stars. For stars with several entries in PASTEL the averaged parameters have been adopted.}
\label{Teff_logg}
\end{figure}

\begin{figure}[t]
\resizebox{\hsize}{!}{\includegraphics{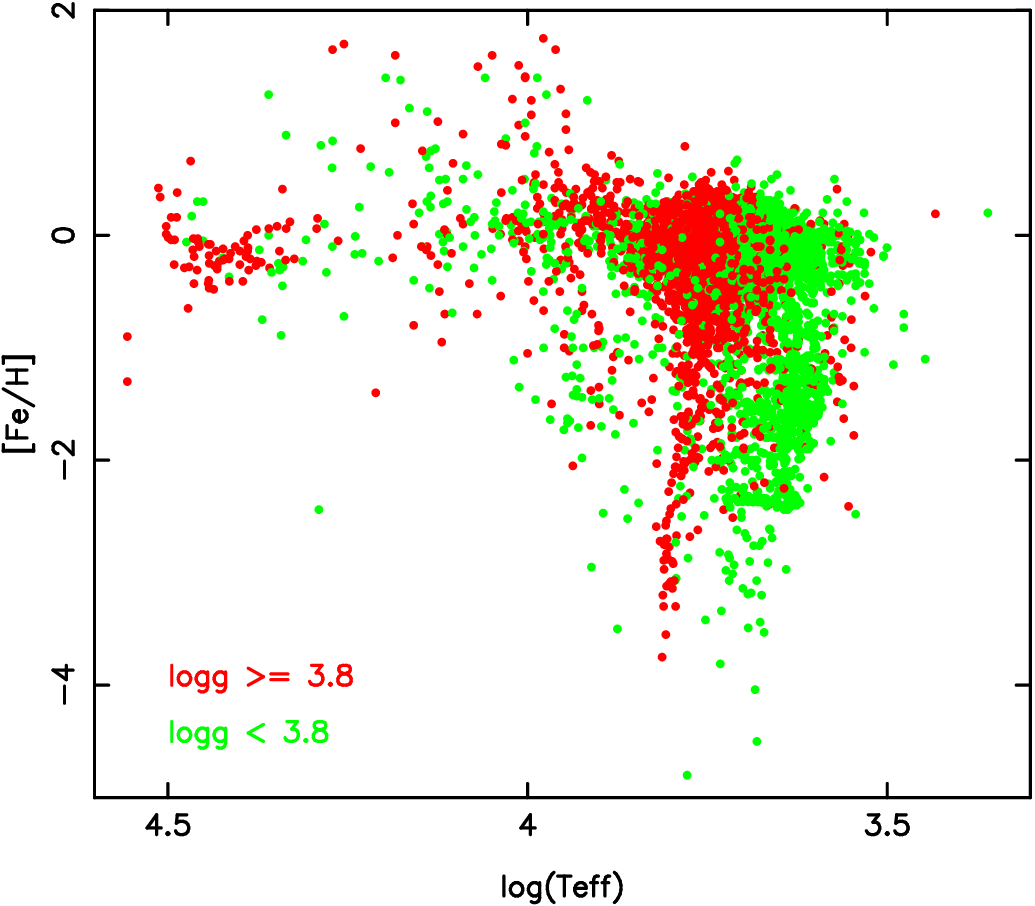}}
\caption{log(\Teff) vs. spectroscopic \FeH for nearly 6000 different stars. Two regimes of gravity are represented, corresponding roughly to giants (green) and dwarfs (red). For stars with several entries in PASTEL, the avaraged parameters have been adopted.}
\label{Teff_FeH}
\end{figure}

There are 14817 entries in PASTEL with the full set of atmospheric parameters \tgm corresponding to 5954 different stars (Figs.~\ref{Teff_logg} and \ref{Teff_FeH}). These numbers were ~6000 and ~3250 respectively in the 1997 \FeH catalogue, with similar values in the 2001 catalogue, but for FGK stars. Despite the improvement of telescopes and spectrographs, there is still  a lack of 
K dwarfs, which are intrinsically faint and more difficult to observe at high resolution 
and high S/N 
than the giants at the same \Teff.  A few M stars have been introduced in the
catalogue, but they are largely underepresented because 
they are difficult to analyse in detail.  

\section {Future evolutions}
In general it is useful to have one single value of \tgm for a given star. It is however quite delicate to simply average determinations from different sources available in PASTEL because they are on different scales, as demonstrated in Sect.~2. Some kind of homogenization has first to be performed, as attempted for instance by \cite{Tay05} and \cite{Twa07} for dwarfs and sub-giants.  We are in the process of building a set of reference stars with homogenized atmospheric parameters, selected from the PASTEL catalogue to cover at best the whole parameter space. We also plan to provide homogenized abundances in PASTEL, following previous work \citep{Sou05}. 
Radial and rotational velocities will also be included in PASTEL.

A useful functionality of  the PASTEL database would be to link it with archives of high resolution spectra  through the Virtual Observatory.  We are working to have soon the PASTEL parameters available for the stellar spectra stored in the archive of the NARVAL spectropolarimeter\footnote{http://tblegacy.bagn.obs-mip.fr/narval.html} attached to the Telescope Bernard Lyot at Pic du Midi. We are also considering such a link with the SOPHIE archive\footnote{http://atlas.obs-hp.fr/sophie/} at Observatoire de Haute-Provence, as well as with the ELODIE archive\footnote{http://atlas.obs-hp.fr/elodie/} \citep{Mou04}.

The content of PASTEL in terms of cluster stars has not changed since the two previous versions of the [Fe/H] catalogue. We are now trying to make up for filling the catalogue with recent extensive studies of open and globuler clusters which have been published since 2001. One of the difficulties with stars in clusters is their correct identification with a name resolved by Simbad (sometimes only charts are available). For open clusters, we have started to work with the WEBDA\footnote{http://www.univie.ac.at/webda/}  team in that sense. 

\section{Conclusion}

We have presented the PASTEL catalogue, which is to date a bibliographical compilation with 30151 entries for 16649 different stars. \Teff determinations  based on various methods are available for all stars. Some 5954 different stars have determinations of the full set of atmospheric parameters \tgm with metallicity based on high resolution, high signal-to-noise spectra. The users are encouraged  to cite the original analysis papers when using the determinations compiled in PASTEL. 

PASTEL offers a useful database for mining stars with known atmospheric parameters, in particular with a high-quality spectroscopic metallicity. The users have to keep in mind that the content of the catalogue is not homogeneous. The stellar content of catalogue, mainly bright stars, is biaised towards stars which are massively studied in peculiar spectroscopic programmes, like solar type stars in planet searches for instance. The \tgm determinations are also not homogeneous and should not be simply averaged for most applications which look into detailed chemical composition of stars.  We have compared metallicity determinations available for the same stars from recent, homogeneous and high quality studies. We have found that in general these studies do not agree well. In some cases the offset can reach 0.07 dex and the dispersion 0.1 dex, larger than the individual errors quoted in the considered studies. The lack of a common \tgm scale results from the variety of observational characteristics, model atmospheres, line data and methods of analysis which are used in spectroscopy. A serious effort should be undertaken to build extensive and homogeneous catalogues of \tgm covering the whole HR diagram and metallicity range. 
 
\begin{acknowledgements}
We warmly thank Ulrike Heiter, Roger Cayrel and Philippe Prugniel for their advices and help in testing the PASTEL database. We made extensive use of the CDS-SIMBAD and NASA-ADS databases and VizieR Service at CDS, and we are  extremely grateful to the staff of these services for maintaining such
valuable resources and for their assistance.
\end{acknowledgements}

\end{document}